# Top-tier and predatory alike? A lexical structure perspective from *the Academy of Management Journal* and *Espacios*

Julián D. Cortés
Universidad del Rosario, Colombia
Fudan University, China
E-mail: julian.cortess@urosario.edu.co

Abstract

This study compares the lexical structure of articles' titles and abstracts' of two extremes in MB (management-business research): the *AMJ (Academy of Management Journal)* –one of its most revered periodicals– and *Espacios* –the one that unveiled a structural problem in Latin-American MB. Results showed significant differences in the median of titles' length and abstracts' readability and diversity as *AMJ* titles' length was longer and abstracts both more diverse and readability-demanding.

Keywords. Readability; Lexical Diversity; Network Analysis; Business, Management and Accounting.

## 1 Introduction

We would celebrate a colleague if his/her research was published in a journal in which more than 15,000 authors have contributed to yield more than 5,800 articles. What a history should that journal be proud of. Indeed, that is the case, as will be shown below.

Lots of things are happening regarding MB (management-business-related research). However, not all are positive, particularly in developing countries. The Winchester Mansion has been used as an analogy for MB: a disorganized dead-end purposeless, although never-ending and expensive infrastructure (Davis, 2015). Moreover, findings have unveiled systematic plagiarism at the heart of leading conferences and business education associations (i.e., The Academy of Management; and the Association to Advance Collegiate Schools of Business – AACSB) (Bedeian et al., 2010; Honig & Bedi, 2012).

Besides, MB output has been strangely brimming in developing countries, raising red flags (Cortés-Sánchez, 2018; Macháček & Srholec, 2021). Remember the journal mentioned at the beginning? (see Table 1 for bibliometric descriptives) A Venezuelan journal with several *predatory features* (Beall, 2020) (i.e., *Espacios*) was the *most desired* journal in the whole Ibero-American region. Researchers from 13 out 22 countries had published at least one article there. The cases of Brazil, Colombia, and Venezuela, were unsettling: 2,890+, 460+, and 100+ articles were published between 1996-2017, respectively (Cortés-Sánchez, 2020a). Additionally, one out of four articles published on innovation in MB in LAC (Latin America and the Caribbean) ended there (Cortés-Sánchez, 2019).

On another discussion, there is a lack of internal disciplinary consistency in MB in LAC, since those topics with higher betweenness among research clusters are those from disciplines seeking consensus (e.g., mathematical models) or sustainability sciences (e.g., sustainable development), and not those from the MB-core (e.g., strategic planning, marketing) (Cortés-Sánchez, 2020b; Cortés, 2021b).

Formulating focused strategies is needed to learn and move toward solution-oriented paths to overcome such disturbances in MB in developing countries (Watts, 2017). For instance, which are the features of reputable MB-journals? As the article's title and abstract are face-value hooks for editors, reviewers, and readers, what are the quality benchmarks for succinctness, clarity, and uniqueness? Are such differences substantially significant between reputable journals and those with *predatory* features? Are both journals publishing diametrically different topics, or are they advocating for topics alike?

Both extremes will be compared here. On the one hand, we consider one of the most reputable journals in MB worldwide: the *AMJ* (*Academy of Management Journal*) (SCImago, 2020). On the other hand, we look at *Espacios*, discontinued from the bibliographic database Scopus as of 2019 (SCImago, 2020). Features such as title and abstract length, abstract clarity and uniqueness (i.e., readability and lexical diversity), and the titles semantic network were computed and compared. Such features are grounded in a mature research agenda (Aleixandre-Benavent et al., 2014; Cortés, 2021a; Didegah & Thelwall, 2013; Haslam et al., 2008; Jamali & Nikzad, 2011; Lewison & Hartley, 2005; Li & Xu, 2019; Méndez et al., 2014; Rostami et al., 2014; Sahragard & Meihami, 2016; Uddin & Khan, 2016; Yitzhaki, 1994, 1997, 2002; Zupic & Čater, 2015).

In that line, this study aims to identify significant differences/similitudes in the lexical structure of article titles and abstracts between the extremes of incidence –or lack of it– in MB. Results could be used by publishers, editors, MB researchers, and the scholarly communication community broadly to understand and apply articles' key features –after considering the disciplinary scope and limitations. After this introduction, the methodology and data section is presented. Then, the results and their discussion with the literature. Finally, the conclusion presents the core findings, limitations, and future agenda.

## 2 Methodology and data

Scopus (2020) was the source of bibliographic data of *Espacios* and *AMJ* articles published in the past ten years (2010-2019). Table 1 presents the bibliometric descriptives. *Espacios'* both output and number of authors for the same time/size sample was over an order of magnitude from that of *AMJ*. In contrast, there is a stationary growth in *AMJ* documents of less than 1% annually. Finally, *AMJ* had over an order of magnitude in terms of citations/document compared to *Espacios*. After confirming the evident article output differences between both journals, 651 articles of *Espacios* were randomly selected for a total sample of 1,302 articles analyzed. The *AMJ* sample was equivalent to ≈39% of the total documents published since 1975 (Scopus, 2020).

Table 1 Sample's bibliometric descriptives

| Journal | Time-span | Documents | Annual growth % | Authors | Authors per document | Citations per document | Sampled articles |
|---|---|---|---|---|---|---|---|
| *Espacios* | 2010-2019 | 5,895 | 39.6 | 15,642 | 2.65 | <1 | 651 |
| *AMJ* | 2019 | 791 | <1 | 1,542 | 1.95 | 88.8 | 651 |

Source: the author based on Scopus (2020) and processed with bibliometrix (Aria & Cuccurullo, 2017).

Table 2 presents the articles' text corpora explored. For both *Espacios* and *AMJ*, titles and abstracts were analyzed. For articles, the number of characters' length and semantic networks were applied. Semantic networks unveil symbols shared and keywords interconnection (Doerfel & Barnett, 1999). To ensure topics' centrality in each article, titles were used over keywords (Nakamura, M; Pendlebury, D; Schnell, J; Szomszor, 2019). Also, titles often contain two of the most relevant indexing keywords for each article (Springer, n.d.; Taylor and Francis, n.d.). Modularity algorithm and betweenness centrality were computed to identify research-topics clusters and topics with a capacity to mediate the flow of information between clusters (Anthonisse, 1971; Blondel et al., 2008).

Readability and lexical diversity were explored in the journals' abstracts. The Flesch-Kincaid grade level (FKGL) index estimates the American school equivalent grade required to comprehend a given text at the first reading (Kincaid et al., 1975). It is used to increase text clarity and exclude unnecessary grammatical complexities. *This sentence has an FKGL of 4.8, which estimates that a person in 7$^{th}$ grade would comprehend it easily.* The Yule's K has been extensively used to assess the lexical richness of texts (Tweedie & Baayen, 1998; Yule, 1944). As Yule's K increase, more words have been repeated, and a lower richness in the text is displayed. A lower Yule's K means a higher lexical richness, therefore fewer words have been repeated in a given text. Harry Potter novels, for instance, have a Yule's K of 69.38-77.12 (Reilly, 2020).

None of the title's length, abstracts' FKGL, nor Yule's K were distributed normally (Shapiro-Wilk tests $p<0.05$) (Shapiro & Wilk, 1965). Consequently, Wilcoxon rank-sum tests were implemented to identify differences between those features for *Espacios* and *AMJ* (Hettmansperger & McKean, 1998).

Table 2 Articles text corpora and techniques applied

| Articles' texts corpora | Techniques applied |
|---|---|
| *Espacios* 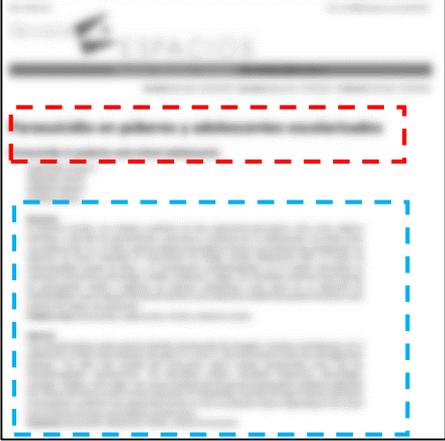 | *Titles - English* <br> Length: number of characters <br> Semantic networks <br> *Abstracts - English* <br> Readability: Flesch-Kincaid grade level $$FKGL = 0.39(w/sen) + 11.8(syll/w) - 15.59$$ Where $w$ refers to the number of words, $sen$ the number of sentences, and $syll$ the number of syllables in a given text. <br><br> Diversity: Yule's K $$K = 10^4 \times \left[ -\frac{1}{N} + \sum_{i=1}^{V} f_v(i,N) \left(\frac{i}{N}\right)^2 \right]$$ Where $N$ refers to the total number of tokens, $V$ to the number of types (i.e., unique tokens), and $f_v(i,N)$ to the number of types occurring $i$ times in a sample of length $N$. |
| *AMJ* 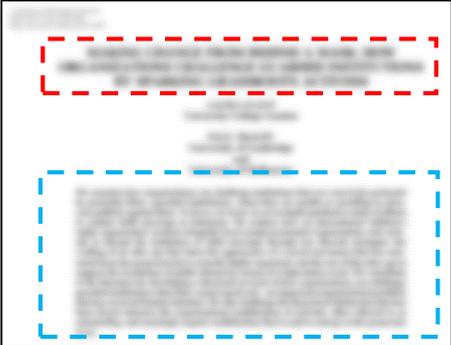 | |

Source: the author based on Scopus (2020), Kincaid et al. (1975), Yule (1944), and processed with Quanteda (Benoit et. al., 2020).

## 3    Results and discussion

Table 3 displays the descriptive statistics for title length (number of characters) and abstracts readability (FKGL) and diversity (Yule's K). Figures 1-3 show density plots for the variables mentioned above. The Wilcoxon rank-sum tests showed that the differences between *Espacios* and *JAM* median titles length ($p=4.52^{-9}$; small effect size $r=0.163$), abstracts readability ($p=6.12^{-15}$; small effect size $r=0.216$) and diversity ($p=2.29^{-111}$; large effect size $r=0.621$) were significant in all cases.

Fifty percent of the titles of the *AMJ* had over 98 characters, compared to those of 89 of *Espacios*. *AMJ* had the shortest title: "*Mediated Sensemaking,*" whereas *Espacios* had the tumultuous:

> "Are our universities really open to the inclusion of the environment in their substantive processes? Analysis of opportunities for curricular environmentalization at the Catholic University of Santiago de Guayaquil and the University of Havana".

It seems more like a paragraph.

In any case, *AMJ* also has admitted titles such as*:*

> *"The multiple pathways through which internal and external corporate social responsibility influence organizational identification and multifoci outcomes: The moderating role of cultural and social orientations*

Behavioral science articles showed no correlation between titles' length and citations (Rostami et al., 2014). Furthermore, it is suggested that titles include a hyphen separating two ideas or at least two different words than the keywords.

Table 3 Title length (number of characters) and abstracts readability (FKGL) and diversity descriptives (Yule's K)

|                    | Min.  | 1st q. | Median | Mean  | 3rd q. | Max.  |
|--------------------|-------|--------|--------|-------|--------|-------|
| *Espacios*         |       |        |        |       |        |       |
| Titles - Length    | 28    | 72     | 89     | 92.31 | 107.5  | 278   |
| Abstract - FKGL    | 9.09  | 14.82  | 17.09  | 17.63 | 19.72  | 58.08 |
| Abstract - Yule's K| 0     | 226.6  | 283.6  | 292.1 | 346.9  | 723.7 |
| *AMJ*              |       |        |        |       |        |       |
| Titles - Length    | 20    | 82     | 98     | 99.07 | 115.5  | 209   |
| Abstract - FKGL    | 11.43 | 16.79  | 18.49  | 18.82 | 20.31  | 51.21 |
| Abstract - Yule's K| 0     | 159.2  | 181.8  | 187.3 | 211.2  | 370.4 |

Source: the author based on Scopus (2020), Kincaid et al. (1975), Yule (1944), and processed with Quanteda (Benoit et. al., 2020).

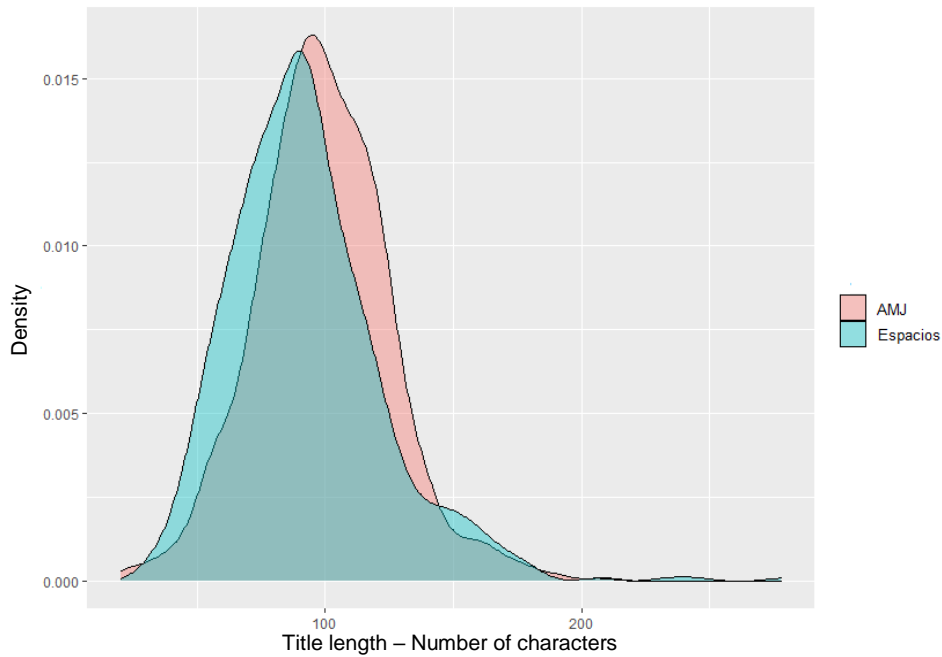

Figure 1 Density plot – Title length: number of characters. Source: the author based on Scopus (2020), Kincaid et al. (1975), Yule (1944), and processed with Quanteda (Benoit et. al., 2020).

Fifty percent of *Espacios*' abstracts had an FKGL over 17.09, whereas for *AMJ,* it was 18.49. A reader would need over 17 years of American schooling to read and understand abstracts at the first reading. Both the most and least readable abstracts were in *Espacios*. Both were under the readability of academic papers, *AMJ* significantly more challenging to read.

The first two sentences of the most readable abstract in *Espacios* are:

> "*It is intended to identify the perception of the bank customers in front of the use of this system. The main limitation to open an account of EM, is the lack of diffusion of this payment mechanism (50%).*"

On the other hand, the first sentence from the least readable is:

> "*This article is the result of an investigation to determine the levels of organizational learning (OL) in administrative processes at the faculty of engineering of university and designing an information management system that supports it.*"

Conversely, the more readable abstract in *AMJ* informs the following:

> "*The importance of an actor's network to his/her private benefits is well explored. Less well understood are the positive externalities of an actor's social capital.*"

The analysis of abstracts from articles published by top research institutions found that such texts are difficult to read (Gazni, 2011). However, there are disciplinary differences: those from mathematics, and pharmacology and toxicology as the more and least readable, respectively. Economics and business were among the top-5 in readability. In general, researchers are increasing their complexity for their writing style and decreasing their overall readability (Plavén-Sigray et al., 2017). Furthermore, that underlines a limitation of readability indices since they do not integrate the reader's previous knowledge to understand an abstract, say, in algebraic topology (Gazni, 2011).

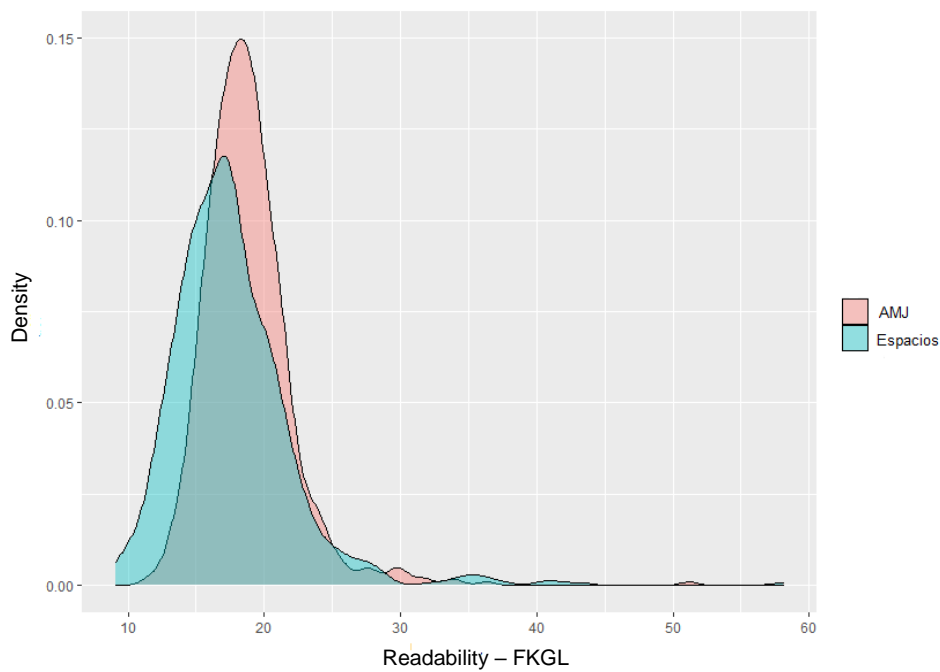

Figure 2 Density plot – Readability: FKGL. Source: the author based on Scopus (2020), Kincaid et al. (1975), Yule (1944), and processed with Quanteda (Benoit et. al., 2020).

Fifty percent of *Espacios'* abstracts have a Yules' K over 283, whereas, for *AMJ* it was 187. Thus, *AMJ* abstracts had way more lexical diversity. The first two sentences of the most diverse abstract in *AMJ* are:

> *"Multiteam systems (i.e., teams of teams) are frequently used to deal with complex and demanding challenges that require several teams' joint efforts. However, achieving effective horizontal coordination across component teams in these systems remains difficult."*

*Espacios'* least diverse abstract reports:

> *"The research revealed the main factors influencing the demand for graduates of higher educational institutions; analyzed the employment of 2010-2017 graduates of the pedagogical institute in Nefu, and summarized the experience of organizing and conducting the pedagogical practice of students."*

Ongoing research dissecting the Yule's K of the mission statement of MB journals, found significant differences between those in the best quartile (i.e., Q1) and the bottom two (i.e., Q3 and Q4) according to the SCImago journal raking (Cortés, 2021a; SCImago, 2020). Those findings support previous research from multiple disciplines that examined the increasing lexical diversity of article titles to address higher communicative attributes and catching readers' attention in leading journals, likewise, the diversification of titles from articles keywords and its association with higher impact (Méndez et al., 2014; Rostami et al., 2014; Yitzhaki, 2002).

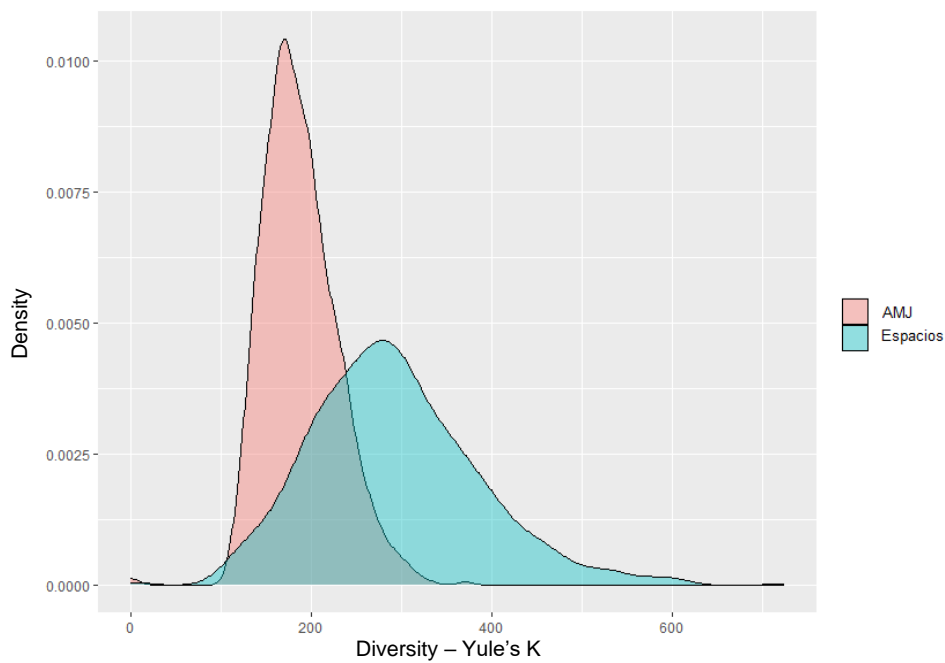

Figure 3 Density plot – Abstracts diversity: Yule's K. Source: the author based on Scopus (2020), Kincaid et al. (1975), Yule (1944), and processed with Quanteda (Benoit et. al., 2020).

Figure 4 shows the semantic networks of the articles' titles. There is a clear-cut between the journals' research clusters. Whereas for *Espacios* the top-3 most crowded clusters were focused on models-process for management systems (green; 9.9% of nodes/title keywords), followed by management education (fuchsia; 9.3%), and employment (gold; 9.18%), for *AMJ* were related to leadership (green; 9.7%), HHRR (fuchsia; 8.8%), and sociologic-related topics (gold; 8.2%).

A vertebral field within MB is strategic management. A co-word analysis conducted for multiple definitions stated that between 1993-2008 its core cluster of terms were *firms, resources, environment,* and *actions* (Ronda-Pupo & Guerras-Martin, 2012). Even though none of the *Espacios* clusters had any relation with those found for strategy, the HHRR cluster of *AMJ* has a clear relation with *resources.*

For *Espacios,* the keyword with the highest betweenness was that of *public* to describe geographic focuses (e.g., *municipal, city, region, Russia*), institutions as norms/rules (e.g., *project, program, policy*), and education actors/activities (e.g., *students, teaching*). For *AMJ* it was that of the *role* of *leadership* and *leaders,* and organizational factors (e.g., *resources, outcomes, management*) and organizational innovation.

Articles' titles-references co-word results in MB in LAC –excluding *Espacios*– found that the co-words with higher betweenness were those related to sustainability sciences (i.e., *sustainable development, environmental impact*), *decision making,* and computer and mathematical-related topics (i.e., *mathematical models, computer simulation, algorithms*) (Cortés-Sánchez, 2020b). Therefore, there is no precise relatedness between those of LAC with *Espacios'.* Likewise, there is more a tangible link with *AMJ* on leadership and decision-making, computer and mathematical-related topics, and organizational (technological) innovation.

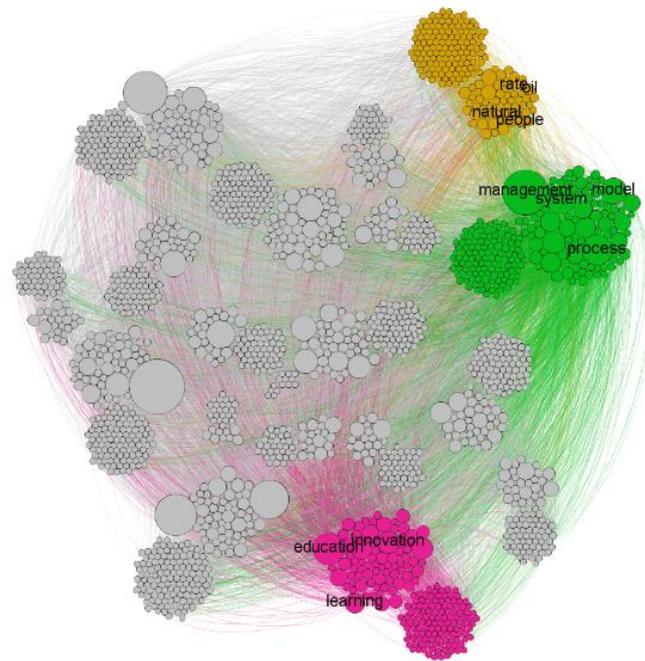

*Espacios*

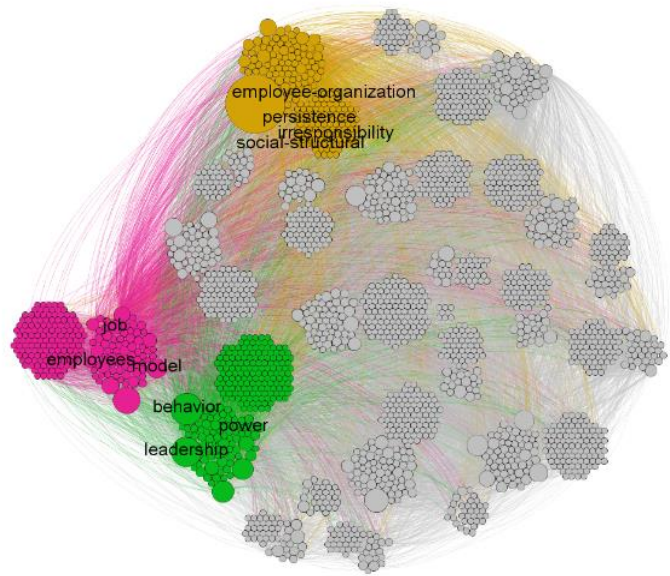

*AMJ*

Figure 4 Titles semantic networks – *Espacios* (top), *AMJ* (bottom). Source: the author based on Scopus (2020). Processed with quanteda, igraph, and Gephi (Bastian et al., 2009; Benoit et. al., 2020; R Core Team, 2014; The igraph core team, 2019). Note: node size proportional to its betweenness score. Labels: top-3 highest degree in each of the top-3 clusters. Network layout algorithm: circle pack (Six & Tollis, 2006).

## 4 Conclusion

The comparison between *Espacios* and *AMJ* showed significant differences in the median of titles' length and abstracts readability and diversity. The titles' length of the *AMJ* was longer, abstracts more readability-demanding, and way more diverse. Such characteristics could be helpful for researchers and editors to consider using an automated proofreader additional to that of a professional translator. Further research could ground these and complementary findings upon a broader perspective on the political economy of science and how socio-economic crises such as that Venezuela is overcoming could boost or shape harmful incentives in research. Also, further research could address more deeply the limitation of using the readability/diversity indices in assessing scholarly communication. As a final note, this article's title length has 99 characters and an abstract with a FKGE of 21.1 and a Yule's K of 309.62.